\documentclass[]{pasj01}

\Received{2022/06/28}
\Accepted{2022/08/30}
 
 \usepackage{lineno}

\begin{document} 

\title{ 
\LETTERLABEL 
A Wide and Deep Exploration of Radio Galaxies with Subaru HSC (WERGS). IX. 
The Most Overdense Region at $z\sim5$ Inhabited by a Massive Radio Galaxy
}

\author{Hisakazu \textsc{Uchiyama}\altaffilmark{1}%
\thanks{Present Address is uchiyama@cosmos.phys.sci.ehime-u.ac.jp}}
\email{uchiyama@cosmos.phys.sci.ehime-u.ac.jp}

\author{Takuji \textsc{Yamashita}\altaffilmark{2,1}}

\author{Tohru \textsc{Nagao}\altaffilmark{1}}

\author{Yoshiaki \textsc{Ono}\altaffilmark{3}}

\author{Jun \textsc{Toshikawa}\altaffilmark{4}}

\author{Kohei \textsc{Ichikawa}\altaffilmark{5,6,7}}

\author{Nozomu \textsc{Kawakatu}\altaffilmark{8}}

\author{Masaru \textsc{Kajisawa}\altaffilmark{1}}

\author{Yoshiki \textsc{Toba}\altaffilmark{2,9,10,1}}

\author{Yoshiki \textsc{Matsuoka}\altaffilmark{1}}


\author{Mariko \textsc{Kubo}\altaffilmark{1}}

\author{Masatoshi \textsc{Imanishi}\altaffilmark{2,11}}

\author{Kei \textsc{Ito}\altaffilmark{12}}

\author{Toshihiro \textsc{Kawaguchi}\altaffilmark{13}}

\author{Chien-Hsiu \textsc{Lee}\altaffilmark{14}}

\author{Tomoki \textsc{Saito}\altaffilmark{15}}

\altaffiltext{1}{Research Center for Space and Cosmic Evolution, Ehime University, 2-5 Bunkyo-cho, Matsuyama, Ehime 790-8577, Japan}
\altaffiltext{2}{National Astronomical Observatory of Japan, Mitaka, Tokyo 181-8588, Japan} 
\altaffiltext{3}{Institute for Cosmic Ray Research, The University of Tokyo, 5-1-5 Kashiwanoha, Kashiwa, Chiba 277-8582, Japan}
\altaffiltext{4}{Department of Physics, University of Bath, Claverton Down, Bath, BA2 7AY, UK}
\altaffiltext{5}{Frontier Research Institute for Interdisciplinary Sciences, Tohoku University, Sendai 980-8578, Japan}
\altaffiltext{6}{Astronomical Institute, Tohoku University, Aramaki, Aoba-ku, Sendai, Miyagi 980-8578, Japan}
\altaffiltext{7}{Max-Planck-Institut f{\"u}r extraterrestrische Physik (MPE), Giessenbachstrasse 1, D-85748 Garching bei M{\"u}unchen, Germany}
\altaffiltext{8}{National Institute of Technology, Kure College, 2-2-11, Agaminami, Kure, Hiroshima 737-8506, Japan}
\altaffiltext{9}{Department of Astronomy, Kyoto University, Kitashirakawa-Oiwake-cho, Sakyo-ku, Kyoto 606-8502, Japan}
\altaffiltext{10}{Academia Sinica Institute of Astronomy and Astrophysics, 11F of Astronomy-Mathematics Building, AS/NTU, No.1, Section 4, Roosevelt Road, Taipei 10617, Taiwan}
\altaffiltext{11}{Department of Astronomical Science, The Graduate University for Advanced Studies, SOKENDAI, Mitaka, Tokyo 181-8588, Japan} 
\altaffiltext{12}{Department of Astronomy, School of Science, The University of Tokyo, 7-3-1 Hongo, Bunkyo-ku, Tokyo, JAPAN, 113-0033}
\altaffiltext{13}{Department of Economics, Management and Information Science, Onomichi City University, Hisayamada 1600-2, Onomichi, Hiroshima 722-8506, Japan} 
\altaffiltext{14}{NSF's National Optical-Infrared Astronomy Research Laboratory, Tucson, AZ 85742, USA} 
\altaffiltext{15}{Nishi-Harima Astronomical Observatory, Centre for Astronomy, University of Hyogo, 407-2 Nichigaichi, Sayo-cho, Sayo, Hyogo 679-5313, Japan}


\maketitle

\begin{abstract}
We report on the galaxy density environment around a high-$z$ radio galaxy (HzRG) at $z=4.72$, HSC J083913.17+011308.1 (HSC J0839+0113), probed  using an $r$-dropout Lyman break galaxy (LBG) sample from the Hyper Suprime-Cam Subaru Strategic Program data.  
We find that  HSC J0839+0113 resides in the outskirt of an overdense region  identified by the $r$-dropout galaxies at  a $4.7\sigma$ significance level.  
The projected distance between HSC J0839+0113 and the peak position of the overdense region is $0.4$ physical Mpc which is shorter than the typical protocluster radius in this epoch. 
According to the  extended Press Schechter and the light cone models, the HSC J0839+0113-hosted overdense region is expected to evolve into a halo $>10^{14}M_\odot$ at $z=0$ with a high probability of $>80$ \%.  
These findings suggest that  HSC J0839+0113 is associated with a protocluster. 
The HSC J0839+0113 rich-system is the most overdense region of LBGs among the known protoclusters with LBGs in the same cosmic epoch. 
\end{abstract} 


\section{Introduction}

High-$z$ radio galaxies (HzRGs) are useful proofs of large-scale cosmic structure,  pointing towards the possible sites of the most overdense regions or rich protoclusters  \citep{Overzier06, Venemans07, Hatch14, Magliocchetti17, Uchiyama21}. 
Study of the most overdense region can provide a strong constraint on the era of the onset of the environmental effects observed in the local universe \citep{Kodama98}. 
However, the number of known HzRGs  at $z>4$ and consequently the number of studies on their environments have been limited. 
This is due to a lack of deep and wide surveys in the radio and optical wavelengths sufficient to capture rare HzRGs and the surrounding galaxies for measuring their environments. 
Only a few of HzRGs have been identified at $z>4$ has been found \citep{Miley08, Saxena19, Yamashita20, Endsley22, Broderick22}.    
It has been confirmed that some of these HzRGs are hosted by overdense regions, such as the 
TN J0924--2201 field at $z=5.2$ \citep{Venemans07, Overzier09}, and COS-87259 field at $z\sim6.8$ \citep{Endsley22b}.

A large survey for radio galaxies, WERGS \citep{Yamashita18, Toba19, Yamashita20, Ichikawa21, Uchiyama21, Uchiyama22}, is now being conducted. 
It is based on the wide ($\sim1000$ deg$^2$) imaging data generated by the Subaru/Hyper Suprime-Cam \citep{Miyazaki12, Miyazaki18} strategic program, HSC-SSP  \citep{Aihara18a}, and the archive catalogs of large radio surveys. 
In this survey,  the HzRG candidates are selected by matching the sources of the  faint images of the radio sky at twenty-cm (FIRST) survey  \citep{Becker95, Helfand15}, with  the  HSC-SSP Lyman break galaxies (LBGs) \citep{Yamashita20, Uchiyama21}. 
The WERGS wide survey can expand the number of HzRGs and could avoid missing rare massive HzRGs. 
Simultaneously, the HSC-SSP data allow us to examine the LBG density environments around the HzRGs. 
One of the WERGS HzRG candidates, HSC J083913.17+011308.1 (hereafter, HSC J0839+0113), has been spectroscopically confirmed at $z = 4.72$  \citep{Yamashita20}. 
Through the rest-frame ultra-violet (UV)-optical multiband photometry, HSC J0839+0113 was found to have a high-stellar mass of $10^{11.4} M_\odot$,  comparable to those of the most massive LBGs at $z\sim5$ \citep{Yamashita20} and that of the HzRG at the same epoch, TN J0924-2201 \citep{Seymour07, Overzier09}.  
Such a massive HzRG is likely to be an ancestor of the brightest cluster galaxy  \citep{West94} and is expected to be hosted by the most overdense region, which can evolve into a rich cluster halo at $z=0$.

In this letter, we report a finding that HSC J0839+0113 is associated with one of the most overdense regions of LBGs at $z\sim5$, based on the HSC-SSP data. 
We assume the following cosmological parameters: 
$\Omega_{M} = 0.3 $, $\Omega_{\Lambda} = 0.7$, $H_{0} = 70 ~ $km s$^{-1}$ Mpc$^{-1}$ $=100 ~ h ~ $km s$^{-1}$ Mpc$^{-1}$.  We use the cModel magnitude \citep{Abazajian04} in the AB system to measure the fluxes.

\section{Samples}

\subsection{HSC J0839+0113}
HSC J0839+0113 was selected from the $r$-dropout galaxies with a FIRST detection and spectroscopically confirmed using Gemini Multi-Object Spectrographs \citep{Yamashita20}. 
The redshift  $z=4.723$ was estimated based on the redshifted Ly$\alpha$ emission line. 
HSC J0839+0113 is the most distant HzRG in the spectroscopically-identified WERGS sample. 
The stellar mass was calculated to be log $M_*/M_\odot =11.43_{-0.46}^{+0.22}$ by fitting the spectral energy distribution (SED) with the multiwavelength data of the optical-to-mid IR bands \citep{Edge13,Cutri14}. 
The rest-frame 1.4GHz radio luminosity is $1.63_{-0.03}^{+0.03} \times 10^{27}$ W Hz$^{-1}$.   
The other spectroscopic and photometric information are available in Table 1 of \citet{Yamashita20}.

\subsection{$r$-dropout galaxies}

To measure the galaxy density around HSC J0839+0113, we use the $r$-dropout galaxies from the HSC-SSP Wide layer data of DR S16A \citep{Aihara18a, Bosch18}. 
The HSC-SSP is a deep-and-wide optical ($g$, $r$, $i$, $z$, and $y$-bands) survey using the HSC with a field-of-view of 1.$^\circ$5 diameter \citep{Aihara18b, Furusawa18, Kawanomoto18, Komiyama18}. 
The  Wide layer data of DR S16A \citep{Aihara18a, Bosch18} 
consists of wide field area of $>200$ deg$^2$ with a median seeing size of $0.\arcsec6 - 0.\arcsec8$. 
Among the spectroscopically-confirmed HzRGs at $z>4$, only HSC J0839+0113  is included in the HSC-SSP area, even after the completion of the HSC-SSP survey. 

The $r$-dropout galaxies are selected as in \citet{Ono18}. 
The HSC-SSP sources are first selected in the $2\times2$ deg$^2$ area centered on HSC J0839+0113 by imposing the HSC-SSP flags summarized in \citet{Ono18}. 
These flags can conservatively mask the regions around bright objects \citep{Coupon18} and bad pixels, and an area of $2.9$ deg$^2$ with uniform depths in all the HSC $grizy$-bands can be selected. 
The $5\sigma$ limiting magnitudes of the $griz$- and $y$-bands for PSF photometry in the effective area are $26.2$, $25.6$, $25.9$, $25.2$, and $24.4$, respectively. 
As a result, the number of  $r$-dropout galaxies is $2043$ in the effective area. 
Certain low-$z$ interlopers such as red galaxies with a prominent Balmer/4000\AA~  break can contaminate the $r$-dropout galaxy sample. 
The contamination rate is estimated to be $\sim30$ \% by \citet{Ono18}.

\section{Density Measurement and Results}

We adopt the fixed aperture method to measure the surface number densities of the $r$-dropout galaxies, following \citet{Toshikawa16}. 
We first uniformly distribute the apertures in a 1 arcmin grid spacing in a sufficiently wide area of $2\times2$ deg$^2$ centered on HSC J0839+0113 to ensure adequate statistics. 
The aperture radius is 1.9 arcmin ($=0.75$ physical Mpc), corresponding to the typical size of a  protocluster at $z\sim5$ with a descendant halo mass $> 10^{14}~ M_{\odot}$ at $z=0$ \citep{Chiang13}. 
Next, we count the number of $r$-dropout galaxies within each aperture.  
If the apertures include parts of the masked regions, the number of  $r$-dropout galaxies is corrected by assuming the average density in the masked regions.
This operation dilutes the possible signals from the real overdense regions, leading to conservative results. 
We do not measure the local density if the center of the aperture overlaps with the masked regions. 
Finally, the overdensity significance, $\Sigma$, is defined as ($N$-$\bar{N}$)/$\sigma$, where $N$ is the number of  $r$-dropout galaxies in each aperture, and $\bar{N}$ and $\sigma$ are the mean and standard deviation of $N$, respectively. 
$\bar{N}$ and $\sigma$ are estimated to be 1.8 and 1.4, respectively. 

\begin{figure}[t]
 \begin{center}
  \includegraphics[width=7.5cm]{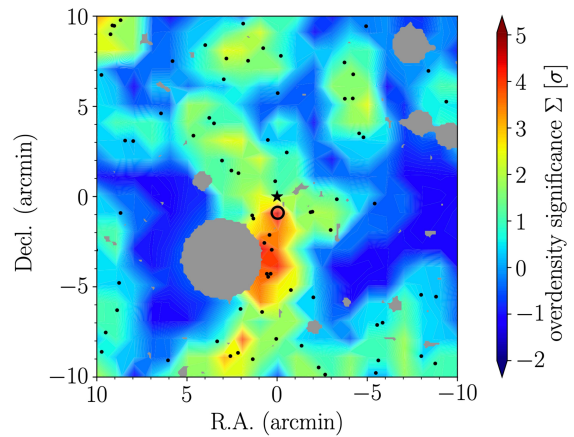} 
 \end{center}
\caption{Overdensity significance map around HSC J0839+0113 (black star). The color contour shows the overdensity significance. The peak position is indicated by the black open circle. The black dots  denote the $r$-dropout galaxies. The gray shades indicate the masked regions.  
}\label{map}
\end{figure}

Figure \ref{map} shows the overdensity significance map around HSC J0839+0113. 
$\Sigma$ at the position of HSC J0839+0113 is $1.6\sigma$. 
We find that HSC J0839+0113 resides in the outskirt of an overdense region with $\Sigma>4\sigma$, south of HSC J0839+0113.   
The peak of $\Sigma$ in this region is $4.7\sigma$. 
The projected distance from HSC J0839+0113 to the peak is 0.4 physical Mpc, which is shorter than the typical radius of a $z\sim5$ protocluster, 0.75 physical Mpc \citep{Chiang13}.  
Thus, HSC J0839+0113 is implied to associate with a protocluster. 

In our density estimation, the average density is used in the masked regions. 
However, even if no $r$-dropout galaxies exist in the masked regions, the  $\Sigma$ peak remains almost the same ($4.5\sigma$). 
In addition, the number density of a $4.7\sigma$ overdense region is expected to be very low at $\sim 1$ deg$^{-2}$ \citep{Toshikawa16}. 
$>4\sigma$ overdense regions occupy $\sim0.3$ \% of the entire sky. 
Thus, such a high density region with a massive rare HzRG  is unlikely to be explained by chance alignment or contamination from nearby sources. 
Besides, we check for possible contamination from low-$z$ groups/clusters. 
If low-$z$ groups/clusters happen to overlap with HSC J0839+0113 in the line-of-sight direction, then a non-negligible number of their red-sequence member galaxies which have red $r-i$ colors could satisfy the $r$-dropout selection, resulting in such large $\Sigma$. 
We find no red-sequence-like tight correlation in a color-magnitude diagram of $r-i$ and $i$ for the $r$-dropout galaxies in the $4.7\sigma$ overdense region. 
We also confirm that there are no ``CAMIRA" clusters \citep{Oguri18} near the $4.7\sigma$ overdense region.  
The CAMIRA clusters are selected by detection of red sequence, based on the HSC Wide layer data which is the same as that used in our study. 
Thus,  it is unlikely that this large $\Sigma$ is produced by low-$z$ group/cluster interlopers. 

\section{Discussion and Summary} 

We found that HSC J0839+0113  was spatially associated with a $4.7\sigma$ overdense region. 
Such overdense regions at $z\sim5$ are expected to evolve into local rich clusters with halo masses  $>10^{14} M_\odot$, with a probability of $\gtrsim80$\% according to the light cone model \citep{Toshikawa16}.  
Following \citet{Uchiyama18,Uchiyama20}, this high probability is also supported by the halo mass of the HzRG as described below. 
\citet{Uchiyama21} performed clustering analysis to estimate the halo masses of the WERGS HzRGs at $z=3.3-4.5$, 
and found that the typical HzRG halo mass is log $M_h/h^{-1}M_\odot=13.02_{-0.14}^{+0.13}$. 
Assuming that HSC J0839+0113 at $z=4.72$ is hosted in a halo with a similar mass,  
 we can evaluate the expected descendant halo mass at $z=0$ using the extended Press Schechter model \citep{Press74, Bower91, Bond91, Lacey93}.  
As a result, the $z=0$ descendant halo mass of the HzRG is estimated to be $4.0_{-0.9}^{+1.1}\times10^{14} h^{-1}M_\odot$ on an average.  
This indicates that the HzRG halo is expected to evolve into a cluster halo at $z=0$ almost  certainly with a probability of $99.8_{-1.9}^{+0.2}$ \%.  
HSC J0839+0113 is expected to be associated with a protocluster. 

It is worth comparing the HSC J0839+0113 system with the spectroscopically-confirmed protoclusters traced by LBGs in the same epoch. 
At $z\sim5$, two protoclusters have been identified with LBGs \citep{Overzier09, Toshikawa20}. 
\citet{Overzier09} found that LBGs with $z\lesssim27.0$ mag are strongly clustered around  HzRG TN J0924-2201  at $z=5.2$.  
In the $3.2\times3.2$ arcmin$^2$ area including TN J0924-2201, the galaxy density is approximately  twice  that of the fields. 
For comparison with the HSC J0839+0113 system, we limit the LBG sample of \citet{Overzier09} to one with $z<25.2$ corresponding to to our $5\sigma$ limiting magnitude. 
We find that the density of the HSC J0839+0113 system is approximately twice that of the TN J0924-2201 field. 
\citet{Toshikawa20} discovered a $z=4.9$ rich protocluster (``D1RD01") without any HzRGs,  by conducting follow-up spectroscopy on  a $\Sigma=4.4\sigma$ overdense region of $r$-dropout galaxies, which was found by use of the Canada-France-Hawaii Telescope Legacy Survey data \citep{Toshikawa16}. 
The D1RD01 region is included in the HSC-SSP Wide layer data of DR S16A. 
The $5\sigma$ limiting magnitude of the HSC-SSP $z$-band in the D1RD01 region is 24.9, which is 0.3 mag shallower than that in our study. 
We select $r$-dropout galaxies in the D1RD01 region based on the same selection criteria used in our study. 
By evaluating the density of the $r$-dropout galaxies with $z<24.9$ within an area of a circle with 1.9 arcmin radius, we find that the density of our HzRG system is 1.2 times higher than that of the D1RD01 region. 
Therefore, the system with which HSC J0839+0113 is associated has a LBG overdensity comparable to or higher than that of the spectroscopically-confirmed protoclusters traced by LBGs in the same epoch.

\begin{figure}
 \begin{center}
  \includegraphics[width=8.0cm]{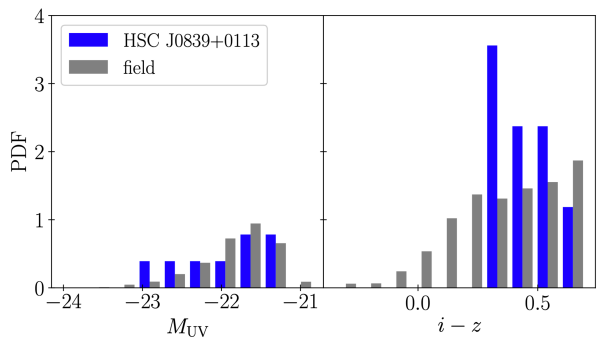} 
 \end{center}
\caption{Probability density function of $M_{\mathrm{UV}}$ (left panel) and $i-z$ (right panel) of the $r$-dropout galaxies in the HSC J0839+0113 overdense region (blue bars) and field (gray bars). 
}\label{color}
\end{figure}

The emergence of the environmental effect, i.e., the dependence of the stellar population or the amount of dust in the galaxy on the ambient density, is already observed in the galaxy overdense regions at $z\sim4$ \citep{Toshikawa16, Ito19}. 
Beyond $z\sim4$, the existence of the environmental effect remains unclear. 
The HSC J0839+0113 system, which is the most overdense region of LBGs in this epoch, can offer insights on the era of the onset of environmental effects. 
The left panel in Figure \ref{color} shows the distribution of the absolute UV magnitudes $M_{\mathrm{UV}}$ of eight $r$-dropout galaxies in the HSC J0839+0113 overdense region, i.e., within 0.75 physical Mpc from the position of the $4.7\sigma$ peak overdensity, compared to that in the field excluding the HSC J0839+0113 overdense region. 
$M_{\mathrm{UV}}$ is estimated as in \citet{Ono18}. 
We find no significant difference in $M_{\mathrm{UV}}$ between the overdense region and the field.   
The $P$-value in the Kolmogorov-Smirnov test is estimated to be 0.51 for the two distributions.  
The right panel in Figure \ref{color} shows the distributions of the $i-z$ colors of the $r$-dropout galaxies in the HSC J0839+0113 overdense region and the field. 
There are no $r$-dropout galaxies with blue color of $i-z\lesssim0.3$ in the HSC J0839+0113 overdense region.  
These two distributions differ from each other, with a $D$-value of $0.45$ and $P$-value of 0.05 in the Kolmogorov-Smirnov test.  
This trend does not change even in the Anderson-Darling test, with an $A^2$-value of 2.6 and a $P$-value of 0.05. 
This result implies that the HSC J0839+0113 overdense region predominantly includes galaxies with an old stellar population and/or large amounts of dust.  
These situations are consistent with the fact that HSC J0839+0113 is hosted by a massive and dusty galaxy \citep{Yamashita20}. 
The galaxies in the HSC J0839+0113 overdense region may have experienced enhanced star formation in an earlier epoch. 
The absence of blue $r$-dropout galaxies suggests the existence of the environmental effect  in the HSC J0839+0113 overdense region. 

\begin{figure}
 \begin{center}
  \includegraphics[width=7.5cm]{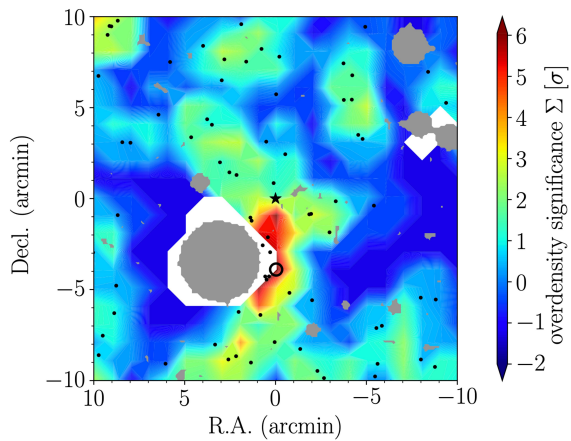} 
 \end{center}
\caption{
Overdensity significance map around HSC J0839+0113 (black star) with a different mask correction. 
The densities of the $r$-dropout galaxies in the masked and unmasked portions of the aperture are equal. 
The white shades indicate the regions where the $>50$ \% of the area is masked in each aperture.  
}\label{map2}
\end{figure}

It is worthwhile to examine the rigidity of the results using a different mask correction method. 
We assume that the densities of the $r$-dropout galaxies in the masked and unmasked portions of an aperture are equal.  
When $>50$ \% of the area is masked in the aperture, we do not measure the local density in the aperture to avoid a considerable uncertainty. 
We find that even in this case, the HSC J0839+0113 overdense region maintains a high-overdensity significance of $\Sigma=6.0\sigma$ (Figure \ref{map2}).  
Hence, HSC J0839+0113 is associated with a protocluster candidate even with this mask correction. 
A large overdense region is clearly visible to the south of the HSC J0839+0113 overdense region. 
The peak significance of the southern overdense region is $\Sigma=6.3\sigma$.  
From the current data, it is not clear whether the HSC J0839+0113 overdense region is separated from the southern overdense region or it is a substructure of a large structure. 
In either case, the HSC J0839+0113 overdense region remains one of the most overdense regions of LBGs at $z\sim5$.

In summary, we examined the galaxy density environment around HSC J0839+0113 with a massive stellar-mass at $z=4.72$.  
We found that HSC J0839+0113 exists in the outskirt of a $4.7\sigma$ overdense region. 
According to the extended Press Schechter  and  light cone models, the HSC J0839+0113  overdense region is expected to evolve into a halo $>10^{14}M_\odot$ at $z=0$, with a high-probability of $>80$ \%.  
The HSC J0839+0113 rich-system is one of the most overdense regions of LBGs compared to the overdensities of the LBGs of two protoclusters in the same epoch. 
The absence of $r$-dropout galaxies with blue colors in the high-density region suggests the existence of the environmental effect in the rich-system. 
So far, the surrounding galaxies are based only on photometric information. 
A wide-area spectroscopic follow-up observation will allow us to scrutinize the overdense region and the underlying Baryon physics.

\begin{ack}
We are deeply grateful to the referee for the helpful comments that assisted in improving the manuscript. 
HU acknowledges support from the JSPS grant 22K14075 and 21H04490. 
TY acknowledges support from the JSPS grant 21K13968 and 20H01939. 
TN acknowledges support from the JSPS grant 20H01949. 

This work is based on data collected at the Subaru Telescope and retrieved from the HSC data archive system, which is operated by Subaru Telescope and Astronomy Data Center at National Astronomical Observatory of Japan. 
The Hyper Suprime-Cam (HSC) collaboration includes the astronomical communities of Japan and Taiwan, and Princeton University. The HSC instrumentation and software were developed by the National Astronomical Observatory of Japan (NAOJ), the Kavli Institute for the Physics and Mathematics of the Universe (Kavli IPMU), the University of Tokyo, the High Energy Accelerator Research Organization (KEK), the Academia Sinica Institute for Astronomy and Astrophysics in Taiwan (ASIAA), and Princeton University. Funding was contributed by the FIRST program from Japanese Cabinet Office, the Ministry of Education, Culture, Sports, Science and Technology (MEXT), the Japan Society for the Promotion of Science (JSPS), Japan Science and Technology Agency (JST), the Toray Science Foundation, NAOJ, Kavli IPMU, KEK, ASIAA, and Princeton University. 

This paper makes use of software developed for the Large Synoptic Survey Telescope. We thank the LSST Project for making their code available as free software at  http://dm.lsst.org

The Pan-STARRS1 Surveys (PS1) have been made possible through contributions of the Institute for Astronomy, the University of Hawaii, the Pan-STARRS Project Office, the Max-Planck Society and its participating institutes, the Max Planck Institute for Astronomy, Heidelberg and the Max Planck Institute for Extraterrestrial Physics, Garching, The Johns Hopkins University, Durham University, the University of Edinburgh, Queen's University Belfast, the Harvard-Smithsonian Center for Astrophysics, the Las Cumbres Observatory Global Telescope Network Incorporated, the National Central University of Taiwan, the Space Telescope Science Institute, the National Aeronautics and Space Administration under Grant No. NNX08AR22G issued through the Planetary Science Division of the NASA Science Mission Directorate, the National Science Foundation under Grant No. AST-1238877, the University of Maryland, and Eotvos Lorand University (ELTE) and the Los Alamos National Laboratory. 

\end{ack}

\end{document}